\documentclass[aps,twocolumn,groupedaddress,nofootinbib,amsmath,amssymb]{revtex4}
\usepackage{dsfont}
\usepackage{color}
\usepackage{bm}
\usepackage[ansinew]{inputenc}
\usepackage[normalem]{ulem}

\begin{document}

\title{Conformally coupled scalar in Lovelock theory}

\author{Eugeny Babichev$^{\dagger}$, Christos Charmousis$^{\dagger}$, Mokhtar Hassaine$^{\ddag}$ and Nicolas Lecoeur$^{\dagger}$}
\affiliation{$^{\dagger}$Universit\'e Paris-Saclay, CNRS/IN2P3,
IJCLab, 91405
Orsay, France,\\
$^{\ddag}$Instituto de Matem\'atica, Universidad de Talca, Casilla
747, Talca, Chile.}

\begin{abstract}
In arbitrary higher dimension, we consider the combination of
Lovelock gravity alongside a scalar-tensor action built out of
higher order operators and Euler densities. The latter action is
constructed in such a way as to ensure conformal invariance for the
scalar field. For the combined version of these theories, we show
the existence of black hole solutions interpreted as stealth
configurations within Lovelock gravity theory. The scalar field
solutions are endowed with an integration constant that may be
identified as a scalar charge. In particular, we show that these
stealth solutions can be extended to include a time-dependent scalar
field despite the underlying theory being non shift-symmetric.
Finally, we present a procedure to obtain a non-conformally
invariant action in even dimensions from the considered theory. For
the target theory, the scalar field is not conformally coupled to
gravity although the scalar field equation itself is conformally
invariant. By means of this procedure, the black hole stealth
configurations are converted into non-stealth black hole solutions,
as discovered recently in four dimensions.

\end{abstract}
 \maketitle

\section{Introduction}

With the given precision of observational data, the theory of
General Relativity (GR) remains unchallenged.
However, given that GR fails to give a self-consistent quantum gravity theory 
and the yet unknown nature of dark energy and dark matter, quite
naturally, the scientific community scrutinizes modified theories of
gravity. One of the simplest, non-trivial yet robust modifications
consists in introducing a scalar field, (non)minimally coupled to
the metric, yielding the so-called scalar tensor theory. The search
for black holes for such theories finds its origin with the
pioneering work of Bocharova, Bronnikov, Melnikov \cite{BBM} and
Bekenstein \cite{Bekenstein:1974sf} who were the first to exhibit a
non trivial, asymptotically flat four-dimensional hairy black hole
with a conformally coupled scalar field. In the Jordan frame,
the action for the scalar field 
is given by the standard kinetic term together with a coupling
between the scalar field $\varphi$ and the scalar curvature $R$,
\begin{eqnarray}
S=\,b_1 \int
d^Dx\sqrt{-g}\left(-\frac{1}{2}(\partial\varphi)^2-\frac{(D-2)}{8(D-1)}R\varphi^2\right)
\label{BBMBaction}
\end{eqnarray}
where $D$ stands for the dimension and $b_1$ is a coupling constant.
The solution to the theory~(\ref{BBMBaction}) in $D=4$  coupled to
the Einstein-Hilbert action, is known as the BBMB solution, with a
metric corresponding to an extremal Reissner-Nordstr\"om spacetime
while the scalar field is shown to blow up at the horizon. Note that
this pathology can be cured by adding a cosmological constant with a
conformally invariant self-interacting
potential~\cite{Martinez:2002ru, Martinez:2005di} and by adding
axionic fields in the case where the $(D-2)-$orthogonal Euclidean
space is a plane \cite{Bardoux:2012tr}. To be more complete, we
mention that black hole solutions with a self-interacting potential
breaking the conformal invariance also exist in four dimensions
\cite{Anabalon:2012tu}, and these latter can be generated by a
certain mapping from the conformal solutions
\cite{Ayon-Beato:2015ada} (see also \cite{Caldarelli:2013gqa}).

In higher dimensions $D> 4$, the extension of the BBMB solution is known, but unfortunately the metric has a naked singularity,  which cannot be removed, unlike the 4-dimensional case~\cite{Klimcik:1993cia}.
More recently, a conformal action generalizing~(\ref{BBMBaction}) was proposed in higher $D$, via
nonminimal couplings of the scalar field with  a four-rank tensor
built out the Riemann tensor and the scalar field.
Such a generalized conformal scalar field coupled to the Einstein or
Lovelock gravity gives rise to an analogue of the BBMB black hole
solutions with (A)dS asymptotics~\cite{Giribet:2014bva}.
The lesson that can be drawn from these studies 
is that the conformal invariance of the action of the scalar field
plays an important role in order to obtain analytical solutions of
the black hole type. One should note however, that the full action
giving rise to the BBMB solution and to its extension in higher
dimensions, is not conformally invariant, since, apart from the
conformally invariant part, the full action contains the
Einstein-Hilbert or Lovelock terms. Strictly speaking, the conformal
symmetry only holds at the level of the equation of motion for the
scalar field. It is then natural to ask whether the conformal
invariance of a part of the action is crucial. Indeed, recently it
has been shown in $D=4$ that this assumption can be relaxed by
requiring only the conformal invariance of the scalar field equation of motion~\cite{Fernandes:2021dsb}.
In this case, two classes of black hole solutions with a regular
scalar field (even in the absence of the cosmological constant)
were found for different fine-tuning of the coupling constants
of the theory \cite{Fernandes:2021dsb}.

In the present paper, we show that the theories constructed in~\cite{Giribet:2014bva} 
admit in addition to the presented solution there, stealth black
hole solutions\footnote{In a scalar-tensor theory, a solution is
called stealth if its spacetime coincides with the one from a pure
metric theory, while having a nontrivial scalar field, which means
that the energy-momentum tensor of the scalar field vanishes
on-shell.}. These are, of the Schwarzschild-(A)dS type, for pure
Einstein gravity, Boulware-Deser spacetimes~\cite{Boulware:1985wk}
in the Einstein-Gauss-Bonnet theory, see \cite{Cai:2001dz}  for the
topological case, and for general  Lovelock theory
~\cite{Wheeler:1985nh}, \cite{Myers:1988ze}.
In all these cases, our solutions have a nontrivial profile for the scalar field with 
an additional constant of integration that may be interpreted as an
independent scalar charge. This scalar charge nevertheless does not
appear in the metric. In other words we have apart from the mass of
the black hole an additional independent charge (not modifying the
metric), therefore the solutions we will describe have neither
primary nor secondary hair. We will refer to this constant simply as
scalar hair.
In addition, 
introducing extra assumptions on the parameters of the action,
stealth configurations defined on the same black hole spacetimes,
albeit with a time-dependent scalar field, can also be constructed.
This result is all the more surprising since the theories under
consideration are not even shift-symmetric. We will see how such a
construction is possible even in the absence of symmetry. Last but
not least, we will present a procedure yielding a non-conformally
invariant 4D action for the scalar field from the generalized
conformal scalar of~\cite{Giribet:2014bva} by performing a singular
limit. The resulting action will lead to a conformally invariant
scalar field equation of Ref.~\cite{Fernandes:2021dsb}. We show that
the singular limit is also compatible at the level of the black hole
solutions, and allows to map the stealth black hole solutions in
higher dimensions to the four-dimensional non stealth black hole
solutions of Ref.~\cite{Fernandes:2021dsb}. Similarly, in higher
even dimensions, (non-stealth) black holes in generalization of the
theory~\cite{Fernandes:2021dsb} are obtained by means of this
singular limit from the stealth black holes.

The plan of the paper is organized as follows. In the next section,
we present  stealth black hole solutions of the
theory~\cite{Giribet:2014bva}.
The extension of this solution to a time-dependent scalar field is
given in Sec. III. The singular limit that allows to construct
non-stealth black hole solutions in a theory with conformal scalar
equation of motion, from stealth solutions of the
theory~\cite{Giribet:2014bva} is explained in Sec. IV. A last
section is devoted to our conclusions.

\section{Stealth black holes with a conformally coupled scalar in Lovelock theory}
In order to be self-contained, we recall the useful formalism and
notations of \cite{Giribet:2014bva} used for the construction of the
most general theory of gravity conformally coupled to a single
scalar field and yielding second-order field equations. This
construction is aimed to generalize the action~(\ref{BBMBaction}).
Indeed, consider a four-rank tensor $S_{\mu\nu}^{\ \ \gamma\delta}$
constructed out of the Riemann curvature tensor $R_{\mu\nu}^{\ \
\gamma\delta}$ and the scalar field, $\phi$,
\begin{align}
S_{\mu\nu}^{\ \ \gamma\delta}  &  =\phi^{2}R_{\mu\nu}^{\ \ \gamma\delta}%
-4\phi\delta_{\lbrack\mu}^{[\gamma}\nabla_{\nu]}\nabla^{\delta]}\phi
+8\delta_{\lbrack\mu}^{[\gamma}\nabla_{\nu]}\phi\nabla^{\delta]}%
\phi-\nonumber\\
&  2\delta_{\lbrack\mu}^{[\gamma}\delta_{\nu]}^{\delta]}\nabla_{\rho}%
\phi\nabla^{\rho}\phi.\label{4tensor}%
\end{align}
where brackets stand for antisymmetrization. One can check to see that this tensor, under a conformal
transformation $g_{\mu\nu}\to \Omega^2 g_{\mu\nu}$ and $\phi\to
\Omega^{-1}\phi$, transforms covariantly, i. e. $S_{\mu\nu}^{\ \
\gamma\delta}\to \Omega^{-4} S_{\mu\nu}^{\ \ \gamma\delta}$. In
arbitrary dimension $D$, the action we will consider is given by
\begin{equation}
S=\int d^{D}x\sqrt{-g}\sum_{k=0}^{\left[  \frac{D-1}{2}\right]
}\frac {1}{2^{k}}\delta^{\left(  k\right)  }\left(  a_{k}R^{\left(
k\right)  } +b_{k}\phi^{D-4k}S^{\left(  k\right)  }\right)
\label{action}
\end{equation}
where $a_k$ and $b_k$ are {\it a priori} arbitrary coupling
constants\footnote{In order to simplify the notations, we will fix
the coupling $a_0=-2\Lambda $ and $a_1=1$.}, where $\delta^{\left(
k\right) }$ is defined as
\[
\delta^{\left(  k\right)  }=(2k)!\ \delta_{\lbrack\alpha_{1}}^{\mu_{1}}%
\delta_{\beta_{1}}^{\nu_{1}}...\delta_{\alpha_{k}}^{\mu_{k}}\delta_{\beta
_{k}]}^{\nu_{k}},
\]
and, where $R^{(k)}$ and $S^{(k)}$ are given by
\begin{equation}
R^{(k)}=\prod\limits_{r=1}^{k}R_{\quad\mu_{r}\nu_{r}}^{\alpha_{r}\beta_{r}%
}\ ,\quad
S^{(k)}=\prod\limits_{r=1}^{k}S_{\quad\mu_{r}\nu_{r}}^{\alpha
_{r}\beta_{r}}\ .
\end{equation}
The $R^{(k)}$ mark Lovelock scalars of increasing rank $k$ ($k=0$
cosmological constant, $k=1$ Einstein-Hilbert, $k=2$ Gauss-Bonnet,
etc.) while $S^{(k)}$, the specific scalar tensor combinations
obtained from  (\ref{4tensor}). It is then easy to see that due to
the covariant transformation of the $4k$-rank $S_{\mu\nu}^{\ \
\gamma\delta}$, the different $b_k$-parts of the action
(\ref{action}) will independently acquire conformal invariance.
Black hole solutions with secondary hair have been obtained for this
theory in Ref.~\cite{Giribet:2014bva}.
We now proceed to show that the theory 
(\ref{action}) admits another class of black hole solutions 
with scalar hair, with an ansatz of the form
\begin{equation}
ds^{2}=-f(r)\
dt^{2}+\frac{dr^{2}}{f(r)}+r^{2}d\Sigma_{D-2,\gamma}^{2},\quad
\phi=\phi(r) \label{ansatz}
\end{equation}
where $d\Sigma_{D-2,\gamma}^{2}$ is the metric of a
($D-2$)-dimensional Euclidean space of  constant curvature
$\gamma\left(D-2\right)\left(D-3\right)$ with $\gamma=(0,\pm1)$.

When the $a_k$-part of the action only contains the Einstein-Hilbert
term with (potentially) a cosmological constant, that is $a_k=0$ for
$k>1$, two different analytic classes of solutions can be found for
the ansatz~(\ref{ansatz}). These two classes correspond to two
different relations between the coupling constants of the action. 
The solutions can be generically given in terms of the metric
functions
\begin{align}
f^{(i)}(r)  &  =\gamma-\frac{M}{r^{D-3}}-
\frac{2\Lambda}{(D-1)(D-2)}r^{2}+\frac{q^{(i)}}{r^{D-2}},
\label{stealthmetricEinsteincase}
\end{align}
dressed with a scalar field given by
\begin{align}
\label{second}
&\phi^{(1)}(r)=\frac{N}{r},\\
&\phi^{(2)}(r) =\frac{N}{r\sigma_{\gamma}\left(c\pm\int
\frac{dr}{\sqrt{f^{(2)}(r)}}\right)}, \label{sfstealth}
\end{align}
where the index $(i)$ denotes the first and the second class of the
solution, and the function $\sigma_{\gamma}$ depends on the topology
of the base manifold
$$
\sigma_{1}(X)=\cosh(X),\quad \sigma_{-1}(X)=\cos(X),\quad
\sigma_{0}(X)=X.
$$
In the above expressions, $M$ is an integration constant
proportional to the mass, while the constant $c$ appearing in the
scalar field, for the second  class of solutions (\ref{sfstealth}),
is the scalar hair. 
The constant $N$ of both scalar fields
(\ref{second})--(\ref{sfstealth}) is fixed in terms of the coupling
constants of the theory through the relation
\begin{align}
\sum\limits_{k=1}^{\left[  \frac{D-1}{2}\right]  }k\
\frac{{b}_{k}}{(D-2k-1)!} \tilde{\gamma}_{(i)}^{k-1}N^{2-2k}  &
=0\text{,}\label{ConstN}
\end{align}
while the coupling of the conformal potential $b_0^{(i)}$ is fixed
in terms of other couplings as
\begin{align}
 \frac{D(D-1)}{(D-1)!}b_0^{(i)}+\sum\limits_{k=1}^{\left[
\frac{D-1}{2}\right]  }\frac{\left(  D\left( D-1\right)
+4\epsilon_k^{(i)}\right){b}_{k}\tilde{\gamma}_{(i)}^{k}}{N^{2k}(D-2k-1)!}
& =0\ , \label{couplingpotential}
\end{align}
with $\epsilon_k^{(1)}=k^2$, $\epsilon_k^{(2)}=k$,
$\tilde{\gamma}_{(1)}=\gamma$ and
$\tilde{\gamma}_{(2)}=\gamma-\delta_{\gamma,0}$. Finally, for both
solutions the constant $q^{(i)}$ appearing in the metric function
(\ref{stealthmetricEinsteincase}) is fixed in terms of the coupling
constants as
\begin{eqnarray}
q^{(i)}=-\frac{b_0^{(i)}}{(D-2)}N^D-\sum\limits_{k=1}^{\left[
\frac{D-1}{2}\right]
}\frac{b_k(D-3)!\tilde{\gamma}_{(i)}^k}{(D-2k-2)!} N^{D-2k}.
\label{qi}
\end{eqnarray}
The first class of solutions  with $i=1$ has $q^{(1)}\not=0$ for
$\gamma\neq 0$ and corresponds to the black hole with secondary hair
found in~\cite{Giribet:2014bva}. For the second class of solutions
for $i=2$, we have $q^{(2)}=0$, and hence the scalar hair solution
can be interpreted as a stealth solution on the Schwarzschild-(A)dS
spacetime, see Eq.~(\ref{stealthmetricEinsteincase}). Importantly, the two classes of spacetimes $i=1,2$ are solutions of distinct theories since $b_0^{(1)}\neq b_0^{(2)}$ as shown by (\ref{couplingpotential}). 

In the general Lovelock case, where $a_k\not=0$ for at least one
$k>1$, similar classes of solutions exist. The scalar field profiles
keep the same form (\ref{second})--(\ref{sfstealth}) and are
subjected to the same conditions (\ref{ConstN}) and
(\ref{couplingpotential}), while  the metric functions $f^{(i)}$
have a different form and are now given by a polynomial equation of
order $\left[ \frac{D-1}{2}\right]$,
\begin{align}
&\sum\limits_{k=0}^{\left[  \frac{D-1}{2}\right] }\frac{a_k(D-1)!}
{(D-2k-1)!}\left(  \frac{\gamma-f^{(i)}\left(  r\right)
}{r^{2}}\right)^{k} =\nonumber\\
&\frac{M(D-1)(D-2)}{ r^{D-1}} -\frac{q^{(i)}(D-1)(D-2)}{r^D}\ ,
\label{poly}
\end{align}
where $M$ is an arbitrary constant related to the mass, and
$q^{(i)}$ are given again by~(\ref{qi}), meaning in particular that
$q^{(2)}=0$. It follows then that the second class of solutions can
be interpreted as stealth black holes of Lovelock theory
(see~\cite{Wheeler:1985nh}, \cite{Myers:1988ze}). In the quadratic
case $a_k=0$ for $k>2$, the real roots of this polynomial can be
easily written down and we have a Boulware-Deser black hole
\cite{Boulware:1985wk} (see \cite{Charmousis:2008kc} for a review)
while for the other cases, the expression for $f$ 
is quite cumbersome, except the case when the polynomial equation
(\ref{poly}) has a single root.
This occurs for the particular choice 
of the coupling constants
$$
a_k=C^{\left[  \frac{D-1}{2}\right]}_k\frac{(D-2k-1)!}{(D-1)!},
$$
which  in odd number of dimensions corresponds to the Chern-Simons point. 
For this particular choice, one can easily express the solution for
metric function in odd dimension as
\begin{eqnarray}
f^{(i)}(r)=\gamma+r^2-\left(\tilde{M}-\frac{\tilde{q}^{(i)}}{r}\right)^{\frac{2}{D-1}},
\label{polyCSodd}
\end{eqnarray}
while in even dimension we have,
\begin{eqnarray}
f^{(i)}(r)=
\gamma+r^2-\left(\frac{\tilde{M}}{r}-\frac{\tilde{q}^{(i)}}{r^2}\right)^{\frac{2}{D-2}},
\label{polyCSeven}
\end{eqnarray}
where we have defined $\tilde{M}=M(D-1)(D-2)$ and
$\tilde{q}^{(i)}=q^{(i)}(D-1)(D-2)$. For the second solution
$\tilde{q}^{(2)}=0$, the spacetime metrics correspond to the black
hole solutions obtained in~\cite{Banados:1993ur}.

\section{Time dependent solutions in theories with no shift symmetry}
As it was originally shown in \cite{Babichev:2013cya}, scalar tensor
theories  with shift symmetry $\phi\to\phi+\mbox{const.}$ may
accommodate black hole solutions with a scalar field that depends
linearly on time. The underlying idea of this feature is that the
field equations only involve derivatives of the scalar field, and
hence its explicit time dependence does not appear at the level of
the field equations.
Here, the action~(\ref{action}) is not shift-symmetric, nevertheless, if
$b_0=b_1=0$ in the action (\ref{action}), the stealth metric
function $f^{(2)}\left(r\right)$ with $q^{(2)}=0$ can be dressed
with a time-dependent scalar field given by {\small
\begin{equation}
\phi\left(t,r\right) =
\exp\left(\hspace{-0.05cm}c+\hspace{-0.05cm}\zeta\,
t+\hspace{-0.1cm}\int\frac{\pm\sqrt{\gamma
f^{(2)}(r)+\zeta^2r^2}/f^{(2)}(r)-1}{r}dr\right), \label{timedep}
\end{equation}}where $c$ and $\zeta$ are arbitrary constants. The emergence of such stealth
solutions in spite of the absence of shift-symmetry in the theory
under consideration can be understood as follows. The vanishing
condition of the energy-momentum tensor of the scalar field can be
schematically written as,
\begin{eqnarray}
\sum_{k\geq 2}b_k\,\phi^{D-2k}{\cal A}_{\mu\nu}^{(k)}=0,
\label{stealthtd}
\end{eqnarray}
where the ${\cal A}_{\mu\nu}^{(k)}$ for $k\geq 2$ only depend on the
derivatives of $\Phi \equiv \log\phi$. One can clearly see that the
above expression is not shift-symmetric, since it involves explicit
dependence on the scalar field, in accord with the fact that the
action is not shift-symmetric. One can verify however that for the
stealth configuration described by the metric function $f^{(2)}(r)$
and the time-dependent scalar field~(\ref{timedep}),
each  ${\cal A}_{\mu\nu}^{(k)}$ vanishes identically, 
and one gets a solution which is effectively shift-symmetric for
$\Phi = \ln\phi$, as highlighted by the form of (\ref{timedep}).

\section{From conformal action to conformal equation}
Here, we present a limiting process in even dimensions which breaks
the conformal symmetry of the scalar field action (\ref{action}) but
still preserving the conformal symmetry of the scalar field
equation. Such an action has been recently proposed in four dimensions
\cite{Fernandes:2021dsb}, and is given in the present notations by
{\small
\begin{eqnarray}
 && S_{\cal{F}}=\int
d^4x\sqrt{-g}\Bigg[R-2\Lambda+b_0\phi^4+b_1\phi^2\left(R+6\frac{(\partial\phi)^2}{\phi^2}\right)\nonumber\\
&&+b_2\left(\log(\phi)\, {\cal
G}-4\frac{G^{\mu\nu}\partial_{\mu}\phi\partial_{\nu}\phi}{\phi^2}-4\frac{\Box\phi
(\partial\phi)^2}{\phi^3}+2\frac{(\partial\phi)^4}{\phi^4}\right)\Bigg],\nonumber\\
\label{fernandesaction}
\end{eqnarray}}where ${\cal G}$ is the Gauss-Bonnet density ${\cal
G}=R^{2}-4R_{\mu\nu}R^{\mu\nu}+R_{\mu\nu\lambda\delta}R^{\mu
\nu\lambda\delta}$. In order to make apparent this limiting process,
let us consider the action~(\ref{action}) in arbitrary dimension $D$
and rewrite it in a similar way, assuming also $a_k=b_k=0$ for
$k>2$, {\small
\begin{widetext}
\begin{eqnarray}
\label{quadraticaction} && S=\int
d^{D}x\sqrt{-g}\Bigg[R-2\Lambda+a_2\,{\cal G}+b_0\,\phi^D+b_1\,\phi^{D-2}\left(R+(D-1)(D-2)\frac{(\partial\phi)^2}{\phi^2}\right)\\
&&+b_2\phi^{D-4}\left( {\cal
G}-4(D-3)(D-4)\frac{G^{\mu\nu}\partial_{\mu}\phi\partial_{\nu}\phi}{\phi^2}-2(D-2)(D-3)(D-4)\frac{\Box\phi
(\partial\phi)^2}{\phi^3}-(D-2)(D-3)(D-4)(D-5)\frac{(\partial\phi)^4}{\phi^4}\right)\Bigg],\nonumber
\end{eqnarray}
\end{widetext}}
\noindent and let us show how the action~(\ref{fernandesaction}) can be obtained from~(\ref{quadraticaction}) by a singular limit. 
This is done by rescaling the coupling constant $b_2\to
\frac{b_2}{D-4}$,  fixing the Gauss-Bonnet coupling
$a_2=-\frac{b_2}{D-4}$, performing a Taylor expansion of
$\phi^{D-4}$ at the neighborhood of $D=4$, i.~e.
$\phi^{D-4}=1+(D-4)\log(\phi)+o(D-4)$, and finally taking the limit
$D\to 4$.
This procedure only works for a non-vanishing Gauss-Bonnet coupling
$a_2$, and hence at the level of the solutions, the limit makes
sense only  for the two classes of solutions presented before in the
Lovelock case, and not for the pure Einstein case. One can verify
that following the above prescription, one recovers two classes of
four-dimensional solutions of the action~(\ref{fernandesaction})
discovered in \cite{Fernandes:2021dsb}, from the solutions
(\ref{second})--(\ref{poly}). In a similar way,  the
higher-dimensional time-dependent stealth solution
of~(\ref{quadraticaction}) with $b_0=b_1=0$, with the scalar given
by~(\ref{timedep}), projects to the time-dependent non-stealth
solution of (\ref{fernandesaction}) with $b_0=b_1=0$ as given
in~\cite{Charmousis:2021npl}.

The same procedure can be easily extended to any even dimension
$D=2p$ with $p\geq 2$ where the Euler density $\delta^{(p)}R^{(p)}$
is a boundary term. Indeed, starting from the action given by
(\ref{action}) with $a_k=b_k=0$ for  $k>p$, one should rescale
$b_p\to \frac{b_p}{D-2p}$, fix the Euler coupling
$a_p=-\frac{b_p}{D-2p}$ and perform a Taylor expansion around
$D=2p$, and finally take the limit $D\to 2p$. The result of this
procedure is an action of a non-conformal scalar field coupled to
Lovelock gravity, yielding however a scalar field equation
{\it{which is}} conformally invariant. Some details of this limiting
procedure are given in the Appendix. Two time-independent solutions
of the resulting action can be read off from
Eqs.~(\ref{second})--(\ref{poly}) by applying them the described
limit. It is worth noting how the procedure works for the
time-dependent solutions in the general
Lovelock theory in even dimension $D=2p>4$. 
Indeed, the energy-momentum tensor (\ref{stealthtd}) must not vanish
since the projected solution yields a non-stealth solution, and must
not depend on the time coordinate $t$ since the metric solution is
time-independent. In fact, one can see that in the considered limit
and after rescaling the couplings, all the ${\cal A}_{\mu\nu}^{(k)}$
of (\ref{stealthtd}) do vanish on the projected solution, except
${\cal A}_{\mu\nu}^{(p)}$, whose time-dependent factor $\phi^{D-2p}$
disappears precisely in the limit $D\to 2p$.

\section{Concluding remarks}
In this paper we presented three main results. First, we showed that
conformally coupled scalar field in Lovelock theory admits a class
of black hole stealth configurations,
Eqs.~(\ref{stealthmetricEinsteincase}), (\ref{sfstealth}) and
(\ref{qi}) with the subscript $i=2$. The metric in this case is
nothing but the Boulware-Deser spacetime \cite{Boulware:1985wk} in
the quadratic case, or its extension for higher Lovelock theory
\cite{Wheeler:1985nh}. The expression of the scalar field contains
the metric function and a constant of integration that may be
interpreted as the scalar charge of the field.

We then demonstrated that in the particular case of the coupling
constants $b_0=b_1=0$ in the action~(\ref{action}), these stealth
configurations can be endowed with a time-dependent  scalar field,
Eq.~(\ref{timedep}).

Finally, a singular limit in even dimension $D=2p$ was presented,
which allows to obtain a non-conformally coupled scalar field,
starting from a conformally coupled invariant scalar field in the
Lovelock gravity. The obtained action includes a direct coupling
between the scalar field and the Euler density of order $p$, which
breaks the conformal symmetry at the level of the action.
Nevertheless for such an action a conformal invariance is kept at
the level of the scalar equation of motion. Within this singular
limit, the black hole stealth configurations (both static and with a
time-dependent scalar) in even dimensions are converted to solution
for black holes with a non-vanishing energy-momentum tensor  of the
scalar field. In~\cite{Charmousis:2012dw}, it was already mentioned
that the non-conformal action (\ref{fernandesaction}) can be
obtained from an alternative Kaluza-Klein (KK) compactification of
the $D-$dimensional Einstein-Gauss-Bonnet theory, while more
recently \cite{Alkac:2022fuc} studied a KK compactification yielding
the same $D=6$ action coupling the scalar field with the cubic Euler
density.

There are yet open questions left for future work concerning
dimensional reduction procedures. In particular, in
Ref.~\cite{Babichev:2022awg}, it was shown that eternal
wormhole-like solutions can also be generated from the
four-dimensional black hole configurations of (\ref{fernandesaction})
by means of a disformal transformation. An interesting question in
the context of our present work is whether these solutions
correspond to some higher dimensional solutions.

\begin{acknowledgments}
We would like to thank Eloy Ay\'on-Beato and Aimeric Coll\'eaux for
useful discussions. We are greatful to ANR project COSQUA for
partially supporting the visit of CC in Talca Chile, where this work
was initiated. The work of MH has been partially supported by
FONDECYT grant 1210889. EB and NL acknowledge support of ANR grant
StronG (ANR-22-CE31-0015-01). The work of NL is supported by the
doctoral program Contrat Doctoral Sp\'ecifique Normalien \'Ecole
Normale Sup\'erieure de Lyon (CDSN ENS Lyon).
\end{acknowledgments}

\appendix
\section{Singular limit}
The action (\ref{action}) can be decomposed as
\begin{equation}
S = S^{(a)}+S^{(b)}
\end{equation}
where $S^{(a)}$ is the pure metric part with coefficients $a_k$, and
$S^{(b)}$ is the scalar-tensor part with coefficients $b_k$ and
enjoying the conformal invariance,
\begin{equation}
S^{(b)}\left[\Omega^2g_{\mu\nu},\Omega^{-1}\phi\right]=S^{(b)}\left[g_{\mu\nu},\phi\right].
\end{equation}
Writing down the vanishing of the variation $\delta S^{(b)}$ under
an infinitesimal transformation $\Omega=1+\epsilon$, one gets the
 identity
\begin{equation}
\label{infinitesimal}
2g^{\mu\nu}\mathcal{E}^{(b)}_{\mu\nu}+\phi\mathcal{E}^{(b)}_\phi = 0
\end{equation}
where
\begin{equation}
\mathcal{E}^{(i)}_{\mu\nu}=\frac{1}{\sqrt{-g}}\frac{\delta
S^{(i)}}{\delta g^{\mu\nu}},\quad
\mathcal{E}^{(i)}_{\phi}=\frac{1}{\sqrt{-g}}\frac{\delta
S^{(i)}}{\delta \phi}.
\end{equation}
Using~(\ref{infinitesimal}) and taking into account that $\mathcal{E}^{(a)}_\phi=0$, one gets that the following combination of the
equations of the full action (noted $\mathcal{E}_{\mu\nu}$, $\mathcal{E}_{\phi}$ with obvious notations) yields a pure geometric constraint:
\begin{equation}
2g^{\mu\nu}\mathcal{E}_{\mu\nu}+\phi\mathcal{E}_\phi =
2g^{\mu\nu}\mathcal{E}^{(a)}_{\mu\nu}=0.
\label{eq:geom} 
\end{equation}

Conversely, it was shown in \cite{Jackiw:2005su, Fernandes:2021dsb}
that an action, such that the combination of field equations given
by the left hand side of (\ref{eq:geom}) is a pure geometric
equation, is not necessarily conformally invariant, but has a scalar
field equation which is conformally invariant. Let us therefore show
that the procedure described in Sec. IV transforms the geometric
equation (\ref{eq:geom}) for $D\geq 2p+1$ in another geometric
equation in the singular limit $D\to 2p$. 
Note that the singular limit procedure does not affect the conformal symmetry of the
Lagrangians $b_k\phi^{D-4k}\delta^{(k)}S^{(k)}$ for $k<p$.
Thus, in order to simplify the presentation, we only focus on actions defined for $D\geq
2p+1$ with $a_k = b_k=0$ for $k\neq p$.  We
therefore have, $\mathcal{E}^{(a)}_{\mu\nu}=G_{\mu\nu}$ and
$\mathcal{E}^{(b)}_{\mu\nu}=-T_{\mu\nu}$ with
\begin{align}
G^\nu_\mu ={}& -\frac{a_p}{2^{p+1}}\delta^{\nu\lambda_1\cdots\lambda_{2p}}_{\mu\rho_1\cdots\rho_{2p}}R^{\rho_1\rho_2}_{\quad\lambda_1\lambda_2}\cdots R^{\rho_{2p-1}\rho_{2p}}_{\quad\lambda_{2p-1}\lambda_{2p}},\nonumber\\
T^\nu_\mu ={}&
\frac{b_p}{2^{p+1}}\phi^{D-4p}\delta^{\nu\lambda_1\cdots\lambda_{2p}}_{\mu\rho_1\cdots\rho_{2p}}S^{\rho_1\rho_2}_{\quad\lambda_1\lambda_2}\cdots
S^{\rho_{2p-1}\rho_{2p}}_{\quad\lambda_{2p-1}\lambda_{2p}},\label{eq:me}
\end{align}
while
\begin{equation}
\mathcal{E}^{(b)}_\phi =
\frac{\left(D-2p\right)b_p}{2^p}\phi^{D-4p-1}\delta^{(p)}S^{(p)},\label{eq:sc}
\end{equation}
see \cite{Giribet:2014bva}. On the other hand, the traces yield
\begin{align*}
G^\nu_\nu ={}& \frac{(2p-D)a_p}{2^{p+1}}\delta^{(p)}R^{(p)},\\\quad
T^\nu_\nu ={}&
-\frac{(2p-D)b_p}{2^{p+1}}\phi^{D-4p}\delta^{(p)}S^{(p)},
\end{align*}
and, hence one gets
\begin{equation}
2g^{\mu\nu}\mathcal{E}_{\mu\nu}+\phi\mathcal{E}_\phi =
\frac{(2p-D)a_p}{2^{p}}\delta^{(p)}R^{(p)}.\label{eq:last}
\end{equation}
It is then easy to see that under the redefinitions $b_p\to\frac{b_p}{D-2p}$ and $a_p\to-\frac{b_p}{D-2p}$, Eqs.~(\ref{eq:sc})--(\ref{eq:last}) have a regular  limit as $D\to 2p$ and that the right hand side of (\ref{eq:last}) is a pure geometric quantity, thus ensuring the conformal symmetry of the scalar field
equation. As for the metric field
equations (\ref{eq:me}), they display a generalized $\delta-$Kronecker
symbol with $2p+1$ indices, which vanish in $D=2p$ and
therefore gives rise to a vanishing factor $(D-2p)$ in dimensional
continuation. This vanishing factor compensates the infinite factor $(D-2p)^{-1}$ from the rescaling of $a_p$ and $b_p$, giving rise to finite metric field equations.
Here,
we have focused on the field equations and proved that the limiting scalar field equation is conformally
invariant. 
Let us now show that the singular limit is also
well-defined at the level of the action.

Up to a global factor $2^{-p}\sqrt{-g}$, the considered Lagrangian
density is
\begin{equation}
\mathcal{L}_p\equiv
a_p\delta^{(p)}R^{(p)}+b_p\phi^{D-4p}\delta^{(p)}S^{(p)},
\end{equation}
and for clarity, we define a function
\begin{align*}
W \equiv {}&{} \phi^{D-4p}\delta^{(p)}S^{(p)} - \phi^{D-2p}\delta^{(p)}R^{(p)}\\
{} = {} & \phi^{D-4p}\delta^{(p)}S^{(p)} - \delta^{(p)}R^{(p)}-
\left(D-2p\right)\left(\log\phi\right) \delta^{(p)}R^{(p)}
\\{}&\hspace{5.8cm}+o\left(D-2p\right).
\end{align*}
Here and in what follows, the notations $o\left(\cdots\right)$ and
$\mathcal{O}\left(\cdots\right)$ have to be understood in the limit
$D\to 2p$. 
The variation of the first two terms in the last expression with respect to the metric are proportional to $T_{\mu\nu}$ and $G_{\mu\nu}$ respectively, see (\ref{eq:me}).
The resulting expressions contain 
a generalized $\delta-$Kronecker  with $2p+1$ indices, which
vanish in $D=2p$. This means that these first two terms are a boundary
term in dimension $D=2p$. Therefore, up to integration by parts, one
has,
\begin{equation*}
\phi^{D-4p}\delta^{(p)}S^{(p)} - \delta^{(p)}R^{(p)} =
\mathcal{O}\left(D-2p\right),
\end{equation*}
and
\begin{equation*}
W = \left(D-2p\right)\tilde{W},
\end{equation*}
with $\tilde{W}$ being regular as $D\to 2p$. 
The Lagrangian $\mathcal{L}_p$ can thus be written as  
\begin{align*} \mathcal{L}_p = & \left(a_p+b_p\right)\delta^{(p)}R^{(p)}\\
 & +b_p\left(D-2p\right)\left[\tilde{W}+\left(\log\phi\right)\delta^{(p)}R^{(p)}+o(1)\right].
\end{align*}
As a consequence, the limiting procedure, namely the rescaling $b_p\to\frac{b_p}{D-2p}$,
$a_p\to -\frac{b_p}{D-2p}$ followed by the limit $D\to 2p$ indeed yields a well defined Lagrangian density,
\begin{equation*}
\mathcal{\tilde{L}}_p=b_p\Bigl[\tilde{W}+\left(\log\phi\right)\delta^{(p)}R^{(p)}\Bigr].
\end{equation*}

\end{document}